\documentclass[a4paper,10pt]{article}

\usepackage[T1]{fontenc}
\usepackage{graphicx}
\usepackage{amsmath}
\usepackage{amssymb}
\usepackage{amsfonts}
\usepackage{bm}
\usepackage{color}

\usepackage{caption}
\usepackage{subcaption}

\usepackage{authblk}

\makeatletter \@addtoreset{equation}{section} \makeatother

\date{}

\def\cL{\mathcal{L}}
\def\cA{\mathcal{A}}
\def\be{\begin{equation}}
\def\ee{\end{equation}}
\def\bea{\begin{eqnarray}}
\def\eea{\end{eqnarray}}

\def\p{\partial}
\newcommand{\mc}[1]{\mathcal{#1}}
\newcommand{\f}[2]{\frac{#1}{#2}}

\title{\boldmath Quantum corrections to the generalized Proca theory via a matter field}

\author{Andr\'e Amado${}^\mu$\thanks{andreamado@df.ufpe.br} , Zahra Haghani${}^\nu$\thanks{z.haghani@du.ac.ir} , Azadeh Mohammadi${}^\lambda$\thanks{a.mohammadi@fisica.ufpb.br}\ \  and Shahab Shahidi${}^\nu$\thanks{s.shahidi@du.ac.ir}\\~\\
\small{${}^\mu$Departamento de F\'isica, Universidade Federal de Pernambuco, 52171-900 Recife, PE, Brazil}\\
\small{${}^\nu$School of Physics, Damghan University, Damghan, 41167-36716, Iran}\\
\small{${}^\lambda$Departamento de F\'isica, Universidade Federal de Campina Grande, Caixa Postal 10071, 58109-970 Campina Grande, PB, Brazil}}

\begin{document}
\unitlength = 1mm
\maketitle

\begin{abstract}
	We study the quantum corrections to the generalized Proca theory via matter loops. We consider two types of interactions, linear and nonlinear in the vector field. Calculating the one-loop correction to the vector field propagator, three- and four-point functions, we show that the non-linear interactions are harmless, although they renormalize the theory. The linear matter-vector field interactions  introduce ghost degrees of freedom to the generalized Proca theory. Treating the theory as an effective theory, we calculate the energy scale up to which the theory remains healthy.
\end{abstract}

\flushbottom

\section{Introduction}
Among all problems in modern cosmology, perhaps the question of why the Universe is in a phase of accelerated expansion at the late time is the most controversial. Many solutions to this question have arisen over the last decades, including the modification of the gravity sector itself, such as $f(R)$ type theories \cite{fR} and massive gravities \cite{massive}. Another approach is to promote the gravitational theory by adding some other degrees of freedom. The simplest possible extension in the second approach is to add one scalar degree of freedom which can describe the accelerated expansion of the Universe, keeping in mind that the theory should remain healthy. Recently, scalar field theories have been proposed in such a way that,
although the action can have higher than second order time derivative terms, the equations of motion remain at most second order.
These scalar fields are now well-known and called Galileons \cite{gali}. In the Minkowski background, these theories have a so-called Galilean symmetry in which the transformation of the scalar field as $\phi\rightarrow\phi+b_\mu x^\mu+c$ leaves the theory invariant. In the curved background however, one should sacrifice this symmetry in face of the necessity of keeping the equations of motion up to second order \cite{gengali}. This theory has been further extended to Multi-Galileon theories \cite{multigali}. Also, many works have been done in order to connect the Galileons to other extended theories \cite{many}. Cosmological implications \cite{cosgali} as well as black hole solutions \cite{blackgali} have been considered in this context.

One of the interesting consequences of the Galileon theories is that the Galileon vertices are not renormalized at the one-loop level \cite{oneloopgali}. One should note that the existence of the Galileon symmetry will prevent the appearance of Galileon symmetry breaking terms at the level of one-loop corrections. However, the Galileon symmetry does not imply that the Galileon vertices are not renormalized. This is the so-called non-renormalization theorem which prevents the theory to become renormalized at one-loop level \cite{non}. This symmetry is known since the appearance of the DGP model \cite{qunDGP}.
Although the non-renormalization theorem guarantees that the theory is not renormalized at one-loop level there is no guarantee that the result will hold if the theory is coupled to a matter field. In fact, the quantum corrections to Galileons due to the matter loops have been calculated in \cite{matterheisenberg}. The author has shown that the matter loops renormalize the Galileon self-interactions, as well as giving rise to ghost instabilities that are harmless to the theory up to specific energy scales.
Therefore, the theory remains healthy within the domain of the effective field theory.

A straightforward way of thinking about the generalization of the idea of Galileons, is to ask whether a vector field theory with such properties exists. One may then think about the promotion of the Maxwell field to a theory with higher derivative terms at the level of the action, while having at most second order time derivatives at the level of the equations of motion. This interesting possibility was considered in \cite{nogo} with the conclusion that there is no such a theory. Assuming that the theory has a $U(1)$ symmetry and at most second order time derivatives at the level of the equations of motion, one can prove that the only possibility is the Maxwell theory itself. However, if one discards the $U(1)$ symmetry, it is possible to consider the Maxwell term plus the mass term for the gauge field, the Proca action, and also generalize it to a self-interacting vector field theory with 3 degrees of freedom \cite{gene}. The generalization of this theory to a covariant theory is straightforward \cite{gene,covgene}. Many works have been done in the context of this generalized Proca theory, including cosmological implications \cite{procos}. Also, there are some generalizations of this theory to include self-interaction terms with the Levi-Civita tensor, etc. \cite{progene}.

One may expect that the non-renormalization theorem does not hold for generalized Proca theory, since the theory has no known field symmetry. In fact, the vector field self-interaction terms can be renormalized at the level of one-loop correction, as discussed in \cite{oneloop}. In that paper the authors have calculated the one-loop effective action of the generalized Proca theory and analytically calculated the one-loop contribution to the two-point function.
Although the generalized Proca theory does not have higher than second order derivatives at the level of the equations of motion, coupling it to a matter field can eventually introduce higher derivatives at quantum level, besides renormalizing the theory itself. In order to understand the relevant domain of applicability of the generalized Proca theory it is important to study the effect of its coupling to matter fields.

It is the scope of the present paper to consider the one-loop corrections to the generalized Proca theory due to the matter loops. We will consider two types of interactions which are linear and non-linear in the vector field. Both interactions will renormalize the generalized Proca theory, but we will discuss that the non-linear interactions are preferable since they do not produce ghost degrees of freedom. When ghost degrees of freedom are introduced in the theory due to quantum corrections, we analyze the energy scale up to which these ghosts remain irrelevant.
After some introductions in section \ref{sec1}, we will make the calculations in section \ref{sec2} and then summarize and discuss the results in the last section.

\section{The Model}\label{sec1}
Let us start with the generalized Proca Lagrangian introduced in \cite{gene}
\begin{align}\label{s1}
    \cL_\text{gen.Proca} = -\frac{1}{4} F_{\mu\nu}F^{\mu\nu} -\frac{1}{2} M^2 \cA^2 + \sum_{n=3}^5 \alpha_n \cL_n,
\end{align}
where
\begin{align}\label{sel1}
	\cL_3 &= f_3(\cA^2)\, \p\cdot\cA,\\
	\cL_4 &= f_4(\cA^2)\, \left[(\p\cdot\cA)^2 + c_1 \p_\rho\cA_\sigma\p^\rho\cA^\sigma-(1+c_1)\p_\rho\cA_\sigma\p^\sigma\cA^\rho\right],\\
	\cL_5 &= f_5(\cA^2)\, \left[(\p\cdot\cA)^3 - 3c_2 (\p\cdot\cA) \p_\rho\cA_\sigma\p^\rho\cA^\sigma-3(1-c_2)(\p\cdot\cA)\p_\rho\cA_\sigma\p^\sigma\cA^\rho\right.\nonumber\\
	&+\left.(2-3c_2)\p_\rho\cA_\sigma\p^\gamma\cA^\rho\p^\sigma\cA_\gamma + 3c_2\p_\rho\cA_\sigma\p^\gamma\cA^\rho\p_\gamma\cA^\sigma\right],
\end{align}
and $c_1$, $c_2$ are two arbitrary dimensionless constants and $f_i$ are some arbitrary functions of $\mc{A}^2\equiv\mc{A}_\mu\mc{A}^\mu$. Also, $F_{\mu\nu}=\partial_\mu\cA_\nu-\partial_\nu\cA_\mu$ is the vector field strength tensor and $\p\cdot\cA$ represents divergence of the vector field. The above Lagrangian has the property that the equations of motion of the vector field are at most second order in time derivatives. It is interesting to note that the terms containing $c_1$ and $c_2$ can be rewritten in terms of the strength tensor $F_{\mu\nu}$.
In the following we will assume that the functions $f_i$ are set to $\cA^2$. This will not change the qualitative behavior of our results while making the calculations simpler \cite{oneloop}.
In the above action, the mass dimension of the vector field $\cA_\mu$ is {\it one}, and $M$ is the mass of the vector field. Also, in the case where $f_i=\cA^2$, the mass dimension of the coupling constants $\alpha_n$ in \eqref{s1} is $2(3-n)$.

\begin{figure}[htp]
	\centering
	\begin{tabular}{cc}
		\begin{subfigure}[a]{0.5\textwidth}
			${}$\qquad\qquad\includegraphics[scale=0.7]{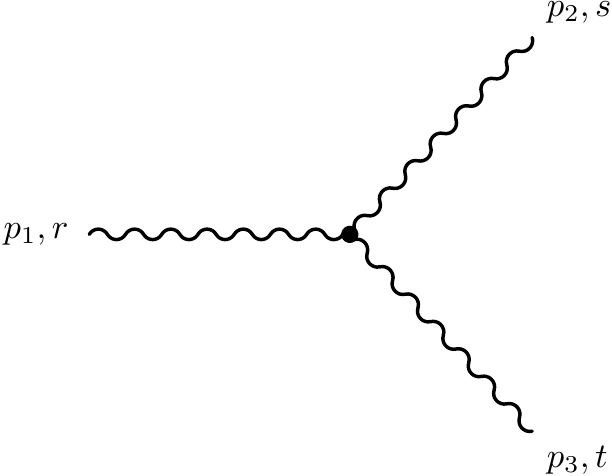}
		\end{subfigure} &
		\begin{subfigure}[a]{0.5\textwidth}
			\includegraphics[scale=0.85]{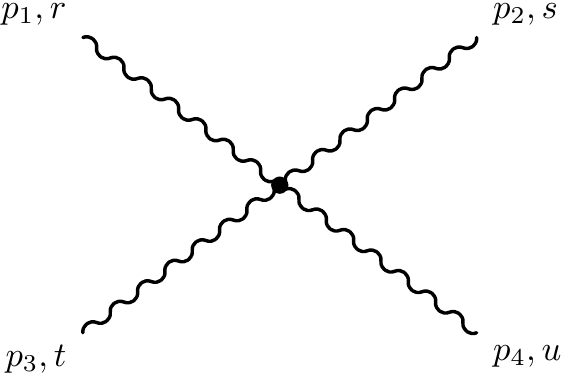}
		\end{subfigure}
	\end{tabular}
	\caption{\footnotesize Feynman diagrams associated to the vector field self-interaction vertices.}
	\label{fig1}
\end{figure}

In this paper, we will consider the one-loop corrections via matter loops to the generalized Proca propagator as well as to the three- and four-point functions of the vector field self-interactions.  Therefore, from now on we will assume $\alpha_5=0$.
The Feynman diagrams associated to these vertices are depicted in figure   \eqref{fig1}.

Let us assume that the matter sector of the theory can be described by a massive scalar field with Lagrangian
\begin{align}
	\cL_{mat} = -\frac{1}{2} \p_\mu\phi\p^\mu\phi - \frac{1}{2} m^2 \phi^2.
\end{align}
The matter energy-momentum tensor can then be obtained as
\begin{align}
	T_{\mu\nu} = \frac{-2}{\sqrt{-g}}\left.\frac{\delta (\cL_{mat}\sqrt{-g})}{\delta g^{\mu\nu}}\right|_{g=\eta} = \p_\mu\phi\p_\nu\phi-\frac{1}{2}\eta_{\mu\nu}\left[(\p\phi)^2+m^2\phi^2\right].
\end{align}
In the following, we will consider four different types of couplings between the matter and the vector fields, which can be separated as
\begin{align}\label{n1}
\cL_{nonlinear} = \frac{1}{M_1^2}\cA^2\, T+\frac{1}{M_2^2}\, \cA^\mu\cA^\nu T_{\mu\nu},
\end{align}
and
\begin{align}\label{n2}
	\cL_{linear} =\frac{1}{M_3^2}\partial_\mu\cA^\mu\, T  + \frac{1}{M_4^2}\, \partial^{(\mu}\cA^{\nu)} T_{\mu\nu},
\end{align}
where $T$ is the trace of the energy-momentum tensor
\begin{align}\label{n3}
T \equiv T_{~\mu}^\mu= -\big[(\p\phi)^2+2m^2\phi^2\big].
\end{align}
The parameters $M_1$, ..., $M_4$ are arbitrary mass scales that characterize the strength of the interactions.
The first terms in equations \eqref{n1} and \eqref{n2}, containing $M_1$ and $M_3$, are conformal couplings between the energy-momentum tensor of the scalar field and the vector field. On the other hand, the terms containing $M_2$ and $M_4$ correspond to disformal couplings.
Also note that the term containing $M_4$ will not contribute to our calculations if the matter energy-momentum tensor is conserved. However, we will keep this term in order to make our calculations general. It is worth mentioning that, if one uses a Maxwell field as a matter field, because the trace of its energy-momentum tensor vanishes, the conformal couplings will not contribute to the calculations.

The following calculations will be carried out on top of the flat Minkowski space. However, one should keep in mind that in the presence of gravity, the above couplings will change the gravitational force in the solar system, through the introduction of a fifth force. In order to screen this extra gravitational force on small scales one should use screening mechanisms such as Vainstein, Chameleon or Symmetron. In our case, because of higher order self-interaction terms for the vector field, the Vainstein mechanism should do the job \cite{scre}.

\begin{figure}[htp]
	\centering
	\begin{tabular}{cc}
		\begin{subfigure}[a]{0.5\textwidth}
			${}$\qquad\qquad\quad\includegraphics[scale=0.9]{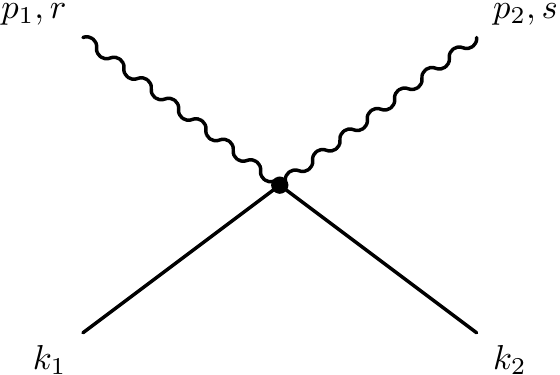}
		\end{subfigure} &
		\begin{subfigure}[a]{0.5\textwidth}
			\includegraphics[scale=0.8]{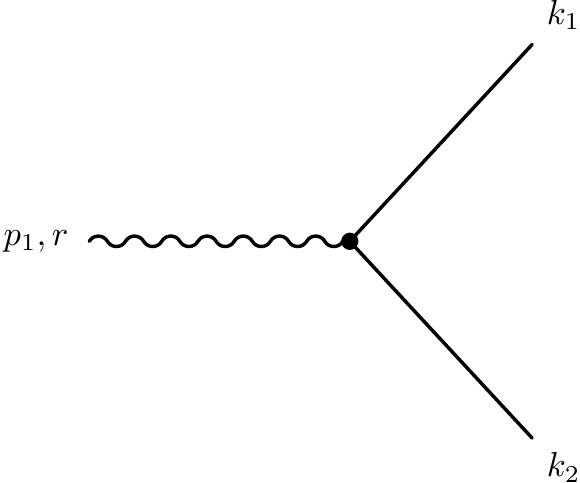}
		\end{subfigure}
	\end{tabular}
	\caption{\footnotesize Feynman diagrams associated with the interaction vertices.}
	\label{fig2}
\end{figure}

From the above interaction terms, one can see that there are three-legged and four-legged vertices. In figure   \eqref{fig2}, we have depicted the corresponding Feynman diagrams.

In order to calculate the one-loop corrections to the generalized Proca theory, we first need to calculate the Feynman propagators of the fields.
The matter field and the vector field propagator can be written as
\begin{align}
	G_\phi = \left\langle\phi(x)\phi(y)\right\rangle = -i\int\frac{\text{d}^4 k}{(2\pi)^4}\frac{e^{ik(x-y)}}{k^2+m^2},
\end{align}
and
\begin{align}
	G_\cA^{\mu\nu}=\langle\cA^\mu(x)\cA^\nu(y)\rangle = -i\int\frac{\text{d}^4 p}{(2\pi)^4}\,\frac{\eta^{\mu\nu}+\frac{p^\mu p^\nu}{M^2}}{p^2+M^2}\,e^{ip(x-y)}.
\end{align}
Now, let us compute the vertex factors associated with the Feynman diagrams in figure   \eqref{fig1}. Let us assume that all momenta go inwards to the vertex. The vertex factors associated with the three- and four-point functions in figure   \eqref{fig1}, can then be obtained respectively as
\begin{align}
V^{3\cA}_{\mu\nu\gamma}(p_1,p_2,p_3) = -2\,\alpha_3\,\left(p_{1\mu}\,g_{\nu\gamma}+p_{2\nu}\,g_{\mu\gamma}+p_{3\gamma}\,g_{\mu\nu}\right),
\end{align}
and
\begin{align}
V&^{4\cA}_{\mu\nu\gamma\delta}(p_1,p_2,p_3,p_4) \nonumber\\=&4\, i\, \alpha_4 \bigg[(
g_{\mu\nu}\,p_{3\gamma}\,p_{4\delta}
+g_{\gamma\delta}\,p_{1\mu}\,p_{2\nu}
+g_{\nu\delta}\,p_{1\mu}\,p_{3\gamma}
+g_{\mu\delta}\,p_{2\nu}\,p_{3\gamma}
+g_{\gamma\nu}\,p_{1\mu}\,p_{4\delta}
+g_{\gamma\mu}\,p_{2\nu}\,p_{4\delta}
)\nonumber\\
+&c_1\big(g_{\mu\nu}\,g_{\gamma\delta}\,(p_1\cdot p_2+p_3\cdot p_4)
+g_{\mu\gamma}\,g_{\nu\delta}\,(p_1\cdot p_3+p_2\cdot p_4)
+g_{\nu\gamma}\,g_{\mu\delta}\,(p_1\cdot p_4+p_2\cdot p_3)\big)
\nonumber\\-&(1+c_1)(g_{\mu\nu}\,p_{3\delta}\,p_{4\gamma}
+g_{\gamma\delta}\,p_{1\nu}\,p_{2\mu}
+g_{\nu\delta}\,p_{1\gamma}\,p_{3\mu}
+g_{\mu\delta}\,p_{2\gamma}\,p_{3\nu}
+g_{\gamma\nu}\,p_{1\delta}\,p_{4\mu}
+g_{\gamma\mu}\,p_{2\delta}\,p_{4\nu})\bigg].
\end{align}
Also, the vertex factors corresponding to the linear and nonlinear interactions can be written respectively as
\begin{align}
V^{2\phi\,1\cA}_\mu(p_1,k_1,k_2)=&-\left(\f{1}{M_3^2}+\f{1}{2M_4^2}\right)(k_1\cdot k_2)p_{1\mu}+m^2\left(\f{2}{M_3^2}+\f{1}{2M_4^2}\right)p_{1\mu}\nonumber\\&+\f{1}{2M_4^2}\big((p_1\cdot k_1)k_{2\mu}+(p_1\cdot k_2)k_{1\mu}\big),
\end{align}
and
\begin{align}
V^{2\phi\,2\cA}_{\mu\nu}(k_1,k_2)
&= i\left[\frac{1}{M_2^2}\, k_{1\mu}k_{2\nu}-\eta_{\mu\nu}(k_1\cdot k_2)\left(\f{1}{M_1^2}+\f{1}{2M_2^2}\right)
+\,m^2\left(\f{2}{M_1^2}+\f{1}{2M_2^2}\right)\eta_{\mu\nu}
\right].
\end{align}

\section{Matter loops}\label{sec2}
Our goal in this paper is to study the quantum corrections to the generalized Proca theory originating from the matter field. In this section we obtain the one-loop quantum corrections to the vector field propagator and self-interaction vertices via the matter loops.

\subsection{One-loop corrections to the propagator}
There are two one-loop contributions of the matter loops to the vector field propagator which can be shown as
\begin{figure}[htp]
\centering
\begin{tabular}{c@{\qquad\qquad}c}
	\includegraphics[scale=0.8]{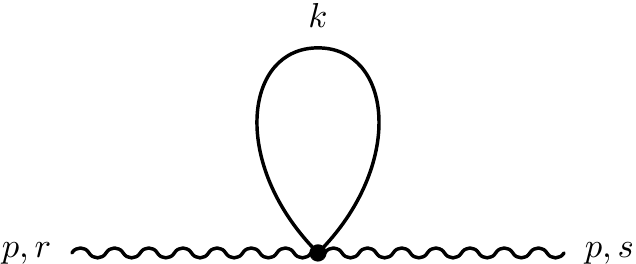}\vspace{0.2cm} & \includegraphics[scale=0.8]{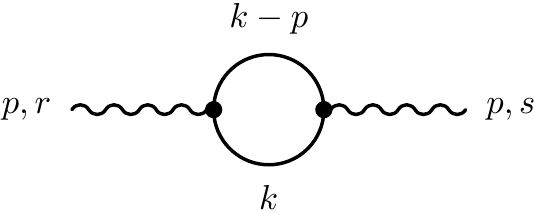}\\
	(a) & (b)
\end{tabular}
\caption{\footnotesize Feynman diagrams associated with one-loop contributions of the matter loops to the vector field propagator.}
\label{propagator}
\end{figure}

The diagram (3-a) gives
\begin{align}
	i\mathcal{M}_{\text{prop,a}} = 2! \cdot \int\frac{\text{d}^4 k}{(2\pi)^4}\ \epsilon^\mu_r(p)\ \frac{-i}{k^2+m^2}\ V^{2\phi\,2\cA}_{\mu\nu}(k,-k)\ \epsilon_s^{\nu*}(p),
\end{align}
which results in
\begin{align}\label{eq32}
	\mathcal{M}_{\text{prop,a}} = 2 m^4\,\epsilon^\mu_r(p) \epsilon^{\nu*}_s(p)\,\delta_{\mu\nu}\left(\frac{1}{M_1^2}+\frac{1}{4M_2^2}\right)J^1_1,
\end{align}
where $r$ and $s$ are polarizations of the vector field. The divergent term in the above equation is encoded in $J^1_1$ which is obtained in the appendix \ref{app1}.

The diagram (3-b) gives
\begin{align}\label{eq33}
	i\mathcal{M}_{\text{prop,b}} &= 2! \cdot 2! \cdot \int\frac{\text{d}^4 k}{(2\pi)^4}\ \epsilon^\mu_r(p)\ \frac{-i}{k^2+m^2}\ 	V^{2\phi\,1\cA}_{\mu}(p,-k,k-p)\ \frac{-i}{(p-k)^2+m^2}\ \nonumber\\
  &\times V^{2\phi\,1\cA}_{\nu}(-p,k,p-k)\ \epsilon_s^{\nu*}(p),
\end{align}
leading to
\begin{align}\label{eq34}
	\mathcal{M}_{\text{prop,b}} = &4 \, \epsilon^\mu_r(p) \epsilon^{\nu*}_s(p) \left[ \frac{m^4}{M_4^4} \frac{p^2}{8}\delta_{\mu\nu} +  \frac{m^4}{M_3^2M_4^2}  \frac{p_\mu p_\nu}{2}  -  \frac{m^4}{M_3^4} p_\mu p_\nu + \frac{p_\mu p_\nu}{M_3^4} \left(\frac{p^4}{4} -\frac{m^2}{2}  p^2 \right) \right] J^1_1.
\end{align}
Therefore, the quantum corrections to the propagator are given by $\mathcal{M}_{\text{prop,a}}+\mathcal{M}_{\text{prop,b}}$.
One can see from equations \eqref{eq32} and \eqref{eq34} that $\mathcal{M}_{\text{prop,a}}$ only renormalizes the mass of the gauge field. On the other hand, $\mathcal{M}_{\text{prop,b}}$ renormalizes the canonical Maxwell kinetic term which scales as $m^4/M_i^4$, where $i=3,4$. These quantum corrections to the propagator will be important when $m\sim ‌M_i$. Moreover, the amplitude $\mathcal{M}_{\text{prop,b}}$ leads to other higher derivative terms which contain ghost degrees of freedom. These terms are obtained from the linear interaction $\partial_\mu A^\mu T$, which can be organized in coordinate space as
\begin{align}
CT_{Prop.}\ni \left\{ \f{1}{M_3^4}(\Box(\partial\cdot\cA))^2\ , \,\f{m^2}{M_3^4}(\partial\cdot\cA)(\Box(\partial\cdot\cA))\, ,\ldots\right\}.
\end{align}
In the above relation for the counterterm, the higher derivative operators will remain negligible provided that $\partial\ll M_3$ \footnote{For the analysis we suppose the mass of the matter field $m$ is smaller than $M_i$ where $i=1,...,4$.}. One should note that other interactions besides the terms corresponding to $M_3$, will not produce ghost degrees of freedom to the propagator.

\subsection{One-loop corrections to the three-point function}
\begin{figure}[htp]
\centering
\begin{tabular}{c@{\qquad\qquad}c}
\includegraphics[scale=0.8]{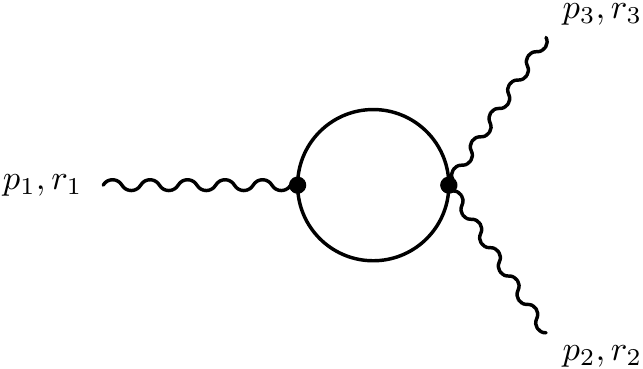}\vspace{0.2cm}& \includegraphics[scale=0.8]{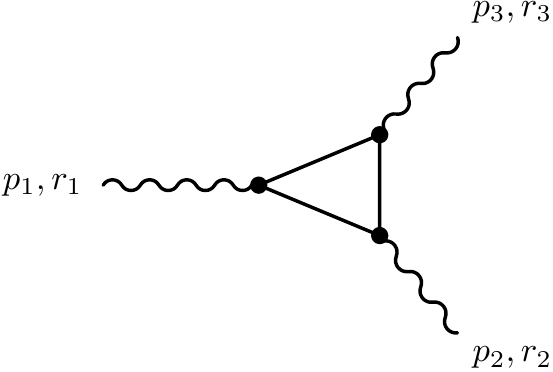}	\\
(a) & (b)
\end{tabular}
\caption{\footnotesize Feynman diagrams associated with one-loop contributions of the matter loops to the three-point function of the generalized Proca theory.}
\label{fig4}
\end{figure}

Now, let us compute the quantum corrections to the three-point function of the generalized Proca theory through matter loops. In this case, we have two diagrams which are relevant to the calculations. These diagrams are depicted in figure \eqref{fig4}. The first diagram which consists of one linear and one non-linear interactions gives
\begin{align}
	i\mathcal{M}_{\text{3pt,a}} &= 3! \cdot 2! \cdot \int\frac{\text{d}^4 k}{(2\pi)^4}\ \epsilon^\mu_{r_1}(p_1)\ \frac{-i}{k^2+m^2}\ V^{2\phi\,1\cA}_{\mu}(p_1,-k,k-p_1)\ \frac{-i}{(k-p_1)^2+m^2}\nonumber \\&\times V^{2\phi\,2\cA}_{\nu\alpha}(k,p_1-k)\ \epsilon_{r_2}^{\nu*}(p_2)\epsilon_{r_3}^{\alpha*}(p_3),
\end{align}
which can be computed as
\begin{align} \label{M3pt1}
	\mathcal{M}_{\text{3pt,a}} &= 3! \cdot 2! \cdot i\ \epsilon_{r_1}^{\mu}(p_1) \epsilon_{r_2}^{\nu*}(p_2)\epsilon_{r_3}^{\alpha*}(p_3) \Bigg\{-\frac{ \delta_{\alpha \nu } \left(-4 m^4-2 m^2 p_1^2+p_1^4\right) p_{1\mu }}{4 M_1^2 M_3^2}+\frac{m^4 p_{1\mu} \delta_{\alpha\nu}}{4 M_1^2 M_4^2}	\nonumber\\&-\frac{p_{1\mu} \left[\left(-3 m^4-2 m^2 p_1^2+p_1^4\right) \delta_{\alpha\nu}-\left(p_1^2-2 m^2\right) p_{1\alpha} p_{1\nu}\right]}{12 M_2^2\,M_3^2}\nonumber\\
&-\frac{m^4 \left(p_{1\nu} \delta_{\alpha\mu}-p_{1\mu} \delta_{\alpha\nu}+p_{1\alpha} \delta_{\mu\nu}\right)}{8 M_2^2 M_4^2}
	\Bigg\} J_1^1.
\end{align}
From the above amplitude, one can see that the term that renormalized the 3-point function is proportional to
$$\f{m^4}{8M_1^2M_2^2M_3^2M_4^2}\Big(2M_2^2(M_3^2+4M_4^2)-M_1^2(M_3^2-2M_4^2)\Big).$$
It is worth mentioning that the interactions proportional to $M_1M_4$ and $M_2M_4$ only renormalize the 3-point functions and do not produce higher derivative terms. So, we conclude that the addition of the interaction $\partial_\mu A^\mu T$ will produce ghost degrees of freedom to the theory. This is the same situation as in the previous section. In the expression (\ref{M3pt1}), there are terms which are not higher derivative interactions and can be written in the coordinate space as
\begin{align}\label{3p1}
CT_{3point}\ni \left\{\f{m^2}{M_3^2M_i^2}(\Box\, \cA^2)(\partial\cdot\cA)\, , \,\f{m^2}{M_2^2M_3^2}(\partial\cdot\cA)^3 \,,\ldots\right\},
\end{align}
where $i=1,2$. These terms have the same form as the terms in the Lagrangian $\mc{L}_5$ with $f_5=1$. One can see that the terms in \eqref{3p1} will renormalize this interaction. The terms which have higher order derivatives and produce ghost degrees of freedom can be written as
\begin{align}\label{3p3}
CT_{3point}\ni \left\{\f{1}{M_3^2M_i^2}(\Box^2\cA^2)(\partial\cdot\cA),\f{1}{M_3^2M_2^2}(\partial\cdot\cA)^2\Box(\partial\cdot\cA),\ldots \right \},
\end{align}
where $i=1,2$. These terms can then be neglected if $\partial^2\ll M_3 M_i \alpha_3$.

The next diagram in figure \eqref{fig4} consists of three linear interactions, and can be written as
\begin{align}\label{3p2}
	i\mathcal{M}_{\text{3pt,b}} =& 3! \cdot 3! \cdot \int\frac{\text{d}^4 k}{(2\pi)^4}\ \epsilon^\mu_{r_1}(p_1) \ \frac{-i}{k^2+m^2}\ V^{2\phi\,1\cA}_{\mu}(p_1,-k,k-p_1)\ \frac{-i}{(k-p_1)^2+m^2}\ \nonumber \\ &\times  V^{2\phi\,1\cA}_{\alpha}(-p_3,p_1-k,p_3+k-p_1)\  \frac{-i}{(p_3+k-p_1)^2+m^2}\ \nonumber \\ &\times V^{2\phi\,1\cA}_{\nu}(-p_2,p_1-k-p_3,k)\ \epsilon_{r_3}^{\alpha*}(p_3)  \epsilon_{r_2}^{\nu*}(p_2).
\end{align}
The computation of the above amplitude is a cumbersome task with  little physical impact. Instead, let us compute this amplitude in a special case of equal momenta where $p_1=p$, $p_2=p$ and $p_3=-2p$. In this case, the amplitude \eqref{3p2} is given by
\begin{align}
	&\mathcal{M}_{\text{3pt,b; eq.m.}}\nonumber\\&\quad = 3! \cdot 3! \cdot \epsilon^\mu_{r_1}(p) \epsilon_{r_2}^{\nu*}(p) \epsilon_{r_3}^{\alpha*}(-2p)  \Bigg\{
	\frac{
		p^2 \left(23 m^2-26 p^2\right) p_{\mu} p_{\nu} p_{\alpha}
	}{M_3^6}+\frac{
	p^2\left(54 m^2 -71 p^2\right) p_{\mu } p_{\nu } p_{\alpha }
}{4 M_3^4 M_4^2}
	\nonumber\\&\quad-\frac{1}{24 M_4^6}
	     p^2 \Big[
	        \left(2 p^4-3 m^4\right) p_{\alpha} \delta_{\mu\nu}
	        +\left(p^4-3 m^4\right) p_{\nu} \delta_{\mu\alpha}
	        +p_{\mu} \left(\left(p^4-3 m^4\right) \delta_{\nu \alpha}
	        +8 p^2 p_{\nu } p_{\alpha }\right)
        \Big]
	\nonumber\\&	\quad
    -\frac{1}{24 M_3^2 M_4^4}
         \bigg[p^2 p_{\mu} \Big(p^2
                \left(12 m^2+13 p^2\right) \delta_{\nu\alpha}
                +16 \left(7 p^2-3 m^2\right) p_{\nu } p_{\alpha }
            \Big)
            \nonumber\\&\quad
                +p^4 \Big(12
                \left(2 m^2+3 p^2\right) p_{\alpha } \delta _{\mu \nu }
                +\left(12 m^2+13 p^2\right) p_{\nu } \delta _{\mu \alpha }
            \Big)
        \bigg]
    \Bigg\}J_1^1.
\end{align}
The interesting fact about this amplitude is that it will not renormalize the 3-point function itself. The terms with minimum number of derivatives correspond to terms like $(\Box \cA^2)(\partial\cdot\cA)$ which scales as $m^4/M_4^6$. These terms will renormalize the self-interaction terms in $\mc{L}_5$ with $f_5=1$. The other terms have five and seven derivatives for three vector fields which will produce ghost degrees of freedom. The terms with five derivatives can be written in the coordinate space as
\begin{align}
CT_{3point}\ni \left\{\left(\f{m^2}{M_3^6}\, , \, \f{m^2}{M_3^4 M_4^2}\, , \, \f{m^2}{M_3^2 M_4^4}\right)(\partial\cdot\cA)^2\Box(\partial\cdot\cA) \, , \, \f{m^2}{M_3^2 M_4^4}(\Box^2 \cA^2)(\partial\cdot\cA)\, ,\ldots \right \}.
\end{align}
It is interesting to note that these terms are the same as \eqref{3p3}. So, this diagram does not introduce extra higher derivative terms to the theory. These terms however, scale as $m^2/M_i^6$ which is smaller than that of equation \eqref{3p3} which scales as $1/M_i^4$. One can see that the above higher derivative terms can be neglected if $\partial\ll M_i \alpha_3$ where $i=3,4$.

The terms containing seven derivatives can be summarized as
\begin{align}
CT_{3point}\ni \Bigg\{\Bigg(\f{1}{M_3^6} \, , \, \f{1}{M_3^4 M_4^2}& \,, \,\f{1}{M_4^6} \, , \,\f{1}{M_3^2 M_4^4}\Bigg)(\partial\cdot\cA)^2\Box^2(\partial\cdot\cA) \,,\nonumber\\ \,&\left(\f{1}{M_4^6} \,, \,\f{1}{M_3^2 M_4^4}\right)(\Box^3 \cA^2)(\partial\cdot\cA) \, ,\ldots \Bigg\}.
\end{align}
These higher derivative terms are new, and have not been produced with the first diagram. For this diagram, one can see that the non-linear interactions do not produce ghost degrees of freedom. In this case, the higher derivative terms can be neglected for $\partial \ll M_i$ where $i=3,4$.

In summary, for the 3-point function, the non-linear interactions will not produce ghost degrees of freedom and just renormalize the 3-point function. Also, if we add the linear interaction terms, we can avoid the ghost considering energies much smaller than the energy scales of the linear interaction terms.

\subsection{One-loop corrections to the four-point function}
\begin{figure}[htp]
\centering
\begin{tabular}{c@{\qquad\qquad}c}
	\includegraphics[scale=0.8]{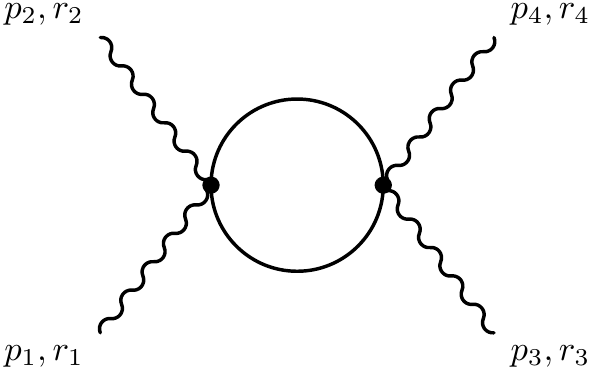}&
	\includegraphics[scale=0.8]{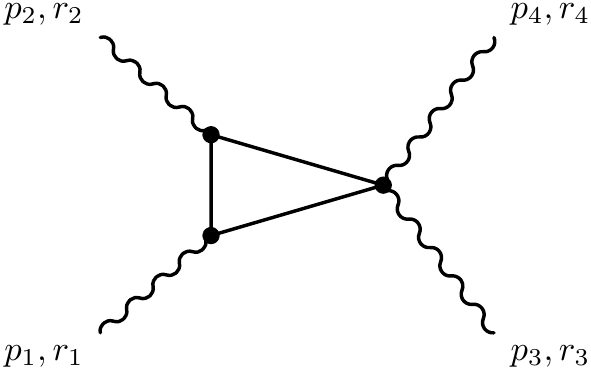}\vspace{0.2cm}\\
	(a) & (b)
\end{tabular}
\\~~\\~~\\
    \includegraphics[scale=0.8]{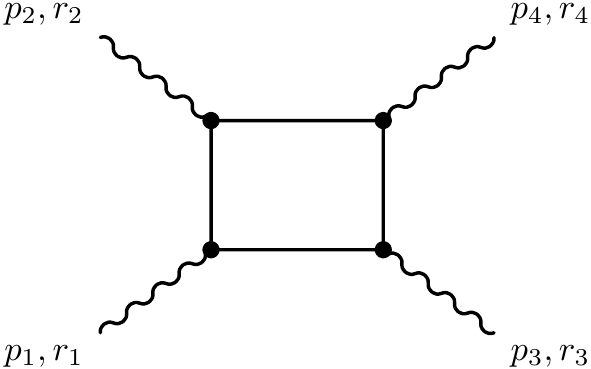}\vspace{0.2cm}\\
    (c)
\caption{\footnotesize Feynman diagrams associated with one-loop contributions of the matter loops to the four-point function of the generalized Proca theory.}
\label{fig5}
\end{figure}

In this section we will compute the one-loop corrections to the 4-point function of the vector field self-interaction terms from scalar matter loops. In this case there are three different diagrams which are depicted in figure \eqref{fig5}.

\noindent
As one can see from the above figure, the diagram (3-a) (the circle diagram) contains two non-linear interactions with the amplitude
\begin{align}
	i\mathcal{M}_{\text{4pt,a}} =& 4! \cdot 2! \cdot \int\frac{\text{d}^4 k}{(2\pi)^4}\ \epsilon^\mu_{r_1}(p_1)\ \epsilon^\nu_{r_2}(p_2)\ \frac{-i}{k^2+m^2}\ V^{2\phi\,2\cA}_{\mu \nu}(-k,k-p_1-p_2)\ \nonumber \\&\times\frac{-i}{(k-p_1-p_2)^2+m^2} V^{2\phi\,2\cA}_{\sigma\rho}(k,p_1+p_2-k)\ \epsilon_{r_3}^{\sigma*}(p_3)\epsilon_{r_4}^{\rho*}(p_4),
\end{align}
after some mathematics, the above amplitude can be written as
\begin{align}
\mathcal{M}_{\text{4pt,a}} &= 4! \cdot 2! \cdot i \epsilon_{r_1}^{\mu}(p_1)\epsilon_{r_2}^{\nu}(p_2)\epsilon_{r_3}^{\alpha*}(p_3)\epsilon_{r_4}^{\beta*}(p_4)\bigg\{\frac{1}{240 M_2^4} \Big[30 m^2 \left(\delta_{\mu\nu} q_{\alpha} q_{\beta} + \delta_{\alpha\beta} q_{\mu} q_{\nu}\right)
\nonumber\\&
-10 m^2 \delta_{(\alpha\beta}\, q_{\mu}\, q_{\nu)}+30 m^4 \left(\delta_{\alpha\nu}\delta_{\beta\mu} - \delta_{\alpha\beta} \delta_{\mu\nu} + \delta_{\alpha\mu}\delta_{\beta\nu}\right)-20 m^2 q_{}^2 \delta_{\alpha\beta} \delta_{\mu\nu}
 \nonumber\\&
-\delta_{\alpha\nu} \left(q_{}^2 q_{\beta} q_{\mu}-\delta_{\beta\mu} \left(10 m^2 q_{}^2+q_{}^4\right)\right)-\delta_{\alpha\mu} \left(q_{}^2 q_{\beta} q_{\nu}-\delta_{\beta\nu} \left(10 m^2 q_{}^2+q_{}^4\right)\right)
 \nonumber\\&
-6 q_{}^2 \delta_{\alpha\beta} q_{\mu} q_{\nu}
-q_{}^2 \delta_{\beta\mu} q_{\alpha} q_{\nu}-q_{}^2 \delta_{\beta\nu} q_{\alpha} q_{\mu}-6 q_{}^2 \delta_{\mu\nu} q_{\alpha} q_{\beta}+6 q_{}^4 \delta_{\alpha\beta} \delta_{\mu\nu}+8 q_{\alpha} q_{\beta} q_{\mu} q_{\nu}\Big]\nonumber\\&
+ \frac{1}{12 M_1^2 M_2^2}\Big[2\delta_{\alpha\beta} \delta_{\mu\nu} \left(q_{}^4-2 m^2 q_{}^2-3 m^4\right)
- \left(q_{}^2-2 m^2\right)\left( \delta_{\mu\nu}\, q_{\alpha} q_{\beta} + \delta_{\alpha\beta}\,q_{\mu} q_{\nu}\right)\Big]
\nonumber\\&+\frac{1}{4 M_1^4}\delta_{\alpha\beta} \delta_{\mu\nu} \left(q_{}^4-2 m^2 q_{}^2-4 m^4\right)\bigg\}J_1^1,
\end{align}
where $q = p_1 + p_2=p_3+p_4$ and
$$\delta_{(\alpha\beta}\, q_{\mu}\, q_{\nu)} =\f{1}{3!}\left( \delta_{\beta\mu} q_{\alpha} q_{\nu}+ \delta_{\beta\nu} q_{\alpha} q_{\mu} + \delta_{\alpha\nu}q_{\beta} q_{\mu} + \delta_{\alpha\mu}q_{\beta} q_{\nu}+ \delta_{\mu\nu} q_{\alpha} q_{\beta} + \delta_{\alpha\beta} q_{\mu} q_{\nu}\right).$$
The above amplitude shows that this diagram does not introduce any higher derivative interactions. However, it introduces some new terms which are not present in the tree-level action. In the first place, there is a self interaction $\cA^4$ which scales as
$$\left(\f{m^4}{M_2^4}\, , \,\f{m^4}{M_1^2M_2^2} \,, \,\f{m^4}{M_1^4}\right).$$
The terms which renormalize the 4-point function of the generalized Proca theory can be summarized as
\begin{align}
CT_{4point}\ni \left \{\f{m^2}{M_2^4}\cA^2(\partial\cdot\cA)^2 \,,\, \left(\f{m^2}{M_2^4},\f{m^2}{M_1^2M_2^2} \,, \,\f{m^2}{M_1^4}\right)\cA^2\Box\,\cA^2,\ldots \right \}.
\end{align}
In the above amplitude, there are some interactions with four derivatives, which can be written as
\begin{align}
CT_{4point}\ni &\left\{\f{1}{M_2^4}(\partial\cdot\cA)^4,\left(\f{1}{M_2^4}\,,\,\f{1}{M_1^2M_2^2}\right)(\partial\cdot\cA)^2\Box\,\cA^2,\right.\nonumber\\
&\left.\left(\f{1}{M_2^4}\,,\,\f{1}{M_1^2M_2^2}\,,\,\f{1}{M_1^4}\right)\cA^2\Box^2\cA^2\,,\ldots\right\}.
\end{align}
These terms are very similar to terms of a Lagrangian with four $\cA$'s and four derivatives, which ought to be $\mc L_6$ with $f_6=1$; i.e. the next order generalized Proca Lagrangian which we did not write in this paper. However, it is well-known that higher order Lagrangians beyond $\mc L_5$ will be total derivatives in four dimensional spacetimes. One should note that the above terms do not produce Ostrogradski ghost to the theory but in general they may produce Boulware-Deser ghost. In fact the generalized Proca Lagrangians \eqref{sel1} are written in such a way that the Boulware-Deser ghost disappears. Any other combinations beyond these Lagrangians will produce a Boulware-Deser ghost degree of freedom.

Now let us consider the next digram (the triangle diagram) which contains two linear and one non-linear interactions. The amplitude of this diagram is
\begin{align}
	i\mathcal{M}_{\text{4pt,b}} &= 4! \cdot 3! \cdot \int\frac{\text{d}^4 k}{(2\pi)^4}\ \epsilon^\mu_{r_1}(p_1)\ \epsilon^\nu_{r_2}(p_2)\ \frac{-i}{k^2+m^2}\ V^{2\phi\,1\cA}_{\mu}(p_1,-k,k-p_1) \ \nonumber \\&\times \frac{-i}{(k-p_1)^2+m^2} V^{2\phi\,1\cA}_{\nu}(p_2,k-p_1-p_2,p_1-k)\ \frac{-i}{(k-p_1-p_2)^2+m^2}\nonumber\\&\times V^{2\phi\,2\cA}_{\sigma \rho}(k,p_1+p_2-k)\ \epsilon_{r_3}^{\sigma*}(p_3)\epsilon_{r_4}^{\rho*}(p_4).
\end{align}
The output of this diagram is very large and we have written the expression for equal momenta case in Appendix \ref{app2}. One can see from the expression \eqref{apb1} that the triangle diagram will renormalize the 4-point function of the theory which scales as
$m^4/(M_2^2M_4^4)$. In the coordinate space, one can write
\begin{align}
CT_{4point}\ni \left \{ \f{m^4}{M_2^2M_4^4}\cA^2\Box\,\cA^2\,,\,\f{m^4}{M_2^2M_4^4}\cA^2(\partial\cdot\cA)^2\,,\ldots \right \}.
\end{align}
As in the previous diagram, the triangle diagram also produces term which should be present in the Lagrangian $\mc{L}_6$ with $f_6=1$. These terms are summarized as
\begin{align}
CT_{4point}\ni \left \{\f{m^2}{M_2^2M_4^4}\cA^2\Box^2\cA^2\,,\,\f{m^2}{M_2^2M_4^4}\Box\,\cA^2(\partial\cdot\cA)^2\,,\,\left(\f{m^2}{M_2^2M_3^4}\,,\,\f{m^2}{M_2^2M_3^2M_4^2}\right)(\partial\cdot\cA)^4\,,\ldots\right \}.
\end{align}
As was mentioned before, although these terms do not introduce Ostrogradski ghost to the theory, they may turn on the Boulware-Deser ghost degree of freedom. Other higher derivative terms can be written in the coordinate representation as
\begin{align}
CT_{4point}\ni &\left \{ \f{1}{M_i^2M_4^4}\cA^2\Box^3\cA^2\,,\,\left(\f{1}{M_i^2 M_j^4}\,,\,\f{1}{M_i^2M_3^2M_4^2}\right) \Box^2\cA^2(\partial\cdot\cA)^2\,,\right.\nonumber\\&\left.
\left(\f{1}{M_2^2M_3^4}\,,\,\f{1}{M_2^2M_3^2M_4^2}\right)\Box(\partial\cdot\cA)^4\,,\ldots\right \},
\end{align}
where $i=1,2$ and $j=3,4$.
One can see that for $\partial^4\ll M_i^2 M_j^4 \alpha_4$, both the Ostrogradski and Boulware-Deser ghost degrees of freedom will be absent from the theory.

The last diagram which may renormalize the 4-point function of the generalized Proca theory contains 4 linear interactions (the square diagram). The amplitude of this diagram can be obtained as
\begin{align}
	i\mathcal{M}_{\text{4pt,c}} =& 4! \cdot 4! \cdot \int\frac{\text{d}^4 k}{(2\pi)^4}\ \epsilon^\mu_{r_1}(p_1)\ \epsilon^\nu_{r_2}(p_2)\ \frac{-i}{k^2+m^2}\epsilon_{r_4}^{\sigma*}(p_4) \epsilon_{r_3}^{\rho*}(p_3) V^{2\phi\,1\cA}_{\mu}(p_1,-k,k-p_1)\ \nonumber \\&\times \frac{-i}{(k-p_1)^2+m^2} V^{2\phi\,1\cA}_{\nu}(p_2,k-p_1-p_2,p_1-k)\ \frac{-i}{(k-p_1-p_2)^2+m^2}\nonumber \\&
	\times V^{2\phi\,1\cA}_{\sigma}(-p_4,k-p_1-p_2+p_4,p_1+p_2-k) \frac{-i}{(k-p_1-p_2+p_4)^2+m^2}\nonumber\\& \times V^{2\phi\,1\cA}_{\rho}(-p_3,k,p_1+p_2-p_4-k).
\end{align}
In Appendix \eqref{app2} we have written the expression for this amplitude in the special case of equal momenta $p_1=p_2=-p_3=-p_4=p$. One can see from equation \eqref{apb2} that this amplitude will not renormalize the 4-point function of the generalized Proca theory itself. The lowest derivative terms has 4 derivatives which makes these term similar to that in $\mc{L}_6$ with $f_6=1$. These terms are as follows
\begin{align}
CT_{4point}\ni\left\{\left(\f{m^4}{M_3^8}\,,\,\f{m^4}{M_3^6M_4^2}\right)(\partial\cdot\cA)^4\,,\,\f{m^4}{M_4^8}\cA^2\Box^2\cA^2\,,\right.\nonumber\\\left.\left(\f{m^4}{M_3^2M_4^6}\,,\,\f{m^4}{M_3^4M_4^4}\right)(\partial\cdot\cA)^2\Box\,\cA^2\,,\ldots\right\}.
\end{align}
These terms scale as $m^4/M_i^8$ with $i=3,4$, which is weaker than their counterparts in the previous diagram by a factor of $m^2/M_i^2$. Other terms have 6 and 8 derivatives which makes the ghost degree of freedom dynamical. Terms with 6 derivatives can be written as
\begin{align}
CT_{4point}\ni&\left\{\f{m^2}{M_4^8}\Big(\cA^2\Box^3\cA^2\,,\,(\partial\cdot\cA)^2\Box^2\cA^2\Big)\,,\,\left(\f{m^2}{M_3^2M_4^6}\,,\,\f{m^4}{M_3^4M_4^4}\right)(\partial\cdot\cA)^2\Box^2\cA^2\right.,\,\nonumber\\&\left.\left(\f{m^2}{M_3^8}\,,\,\f{m^2}{M_3^6M_4^2}\right)\Box(\partial\cdot\cA)^4\,,\ldots\right\}.
\end{align}
These terms scale as $m^2/M_i^8$ which is again weaker than the previous diagram by a factor of $m^2/M_i^2$. Finally, terms with 8 derivatives can be summarized as
\begin{align}
CT_{4point}\ni \left \{\f{1}{M_4^8}\cA^2\Box^4\cA^2\,,\,\left(\f{1}{M_3^8}\,,\,\f{1}{M_3^6M_4^2}\,,\,\f{1}{M_4^8}\right)\Big(\Box^2(\partial\cdot\cA)^4\,,\,(\partial\cdot\cA)^2\Box^3\cA^2\Big)\,,\ldots \right \},
\end{align}
which scales as $1/M_i^8$. One can see that all the above terms can cause ghost degrees of freedom to the theory. The first set can turn on the Boulware-Deser ghost and the rest will make the Ostrogradski ghost dynamical. However, for scales $\partial^6\ll M_i^8 \alpha_4$ where $i=3,4$, these ghost degrees of freedom become non-dynamical.

\section{Conclusion and final remarks}
In this paper we have considered the quantum corrections to the generalized Proca action due to an external matter field. We have calculated the one-loop correction to the gauge field propagator as well as three- and four-point functions, considering two different types of interactions with matter field. The ones considered in this paper are linear and nonlinear in the gauge field.

Calculating the one-loop quantum corrections to the vector field propagator, we have shown that the diagram (3-a) contributes only to the mass term correction, while the diagram (3-b) renormalizes the canonical Maxwell kinetic term as well as introducing higher derivative terms which give rise to ghost degrees of freedom. The higher derivative terms remain irrelevant as long as we have $\partial\ll M_3$. Two diagrams contribute to the one-loop correction of the three-point function. Besides renormalizing the three-point function, they create some terms with the same form of terms appearing in $\cL_5$ with $f_5=1$ and higher derivative terms. To suppress the ghost degrees of freedom it is enough that $\partial\ll M_i \alpha_3$ where $i=1,\dots,4$. For the one-loop correction to the four-point function, three diagrams contribute. In this case, there appear terms renormalizing the four-point function besides the ones with the same shape as terms appearing in $\cL_6$ with $f_6=1$. Although these terms do not create any ghost, the diagrams include some other terms which produce higher derivatives with ghost problem. To avoid the problem within the validity of the effective theory, we can consider the constraint $\partial^6\ll M_i^8 \alpha_4$ where $i=1,\dots,4$.

In the special case of zero momenta one can see that the contribution to the one-loop corrections come from two diagrams only, diagrams (3-a)
and (5-a). Both diagrams arise from the non-linear interactions. It is interesting to notice that the three-point function does not receive any correction in this case.
In general, in the absence of linear interactions the three-point function receives no corrections. Non-zero momentum corrections are generated by diagrams containing either combinations of linear and non-linear interactions or linear interactions only. This way restricting the interactions to non-linear in the vector field ones is one possible way of guaranteeing the theory is ghost free.
Analyzing all one-loop correction diagrams we can see that under the condition $\partial\ll M_i$ for all $i=1,\dots,4$ the ghost degrees of freedom stay non-dynamical, which guarantees that the theory is healthy within this domain.

One can infer from the computations in this paper that the quantum corrections to the five-point function of this theory should be qualitatively equivalent to the results of the four-point function calculations. As a result, one can expect there should appear ghost degrees of freedom due to the linear interactions and also terms which are similar to the would be $\cL_7$ with $f_7=1$.

\appendix
\section{Computation of the loop integrals}\label{app1}
In the calculations of this paper there appear integrals of the form
\begin{align}
J_n^l=\frac{1}{m^4}\int\frac{\text{d}^4 k}{(2\pi)^4}\ \frac{k^{2l}}{(k^2+m^2)^n},
\end{align}
where $l=0,1,2,...$ and $n=1,2,...$\,. Using dimensional regularization one can obtain
\begin{align}
J_n^l= \begin{cases}
\frac{i m^{2(l-n)}}{(4\pi)^2}\frac{\Gamma\left(l+2\right)\,\Gamma\left(n-l-2\right)}{\Gamma(n)}, & \quad n-l-2 > 0\\
(-1)^{l-n}m^{2(l-n)}\frac{\Gamma(l+2)}{\Gamma(n)\Gamma(l-n+3)}\left[J^1_1-\frac{i}{(4\pi)^2}\sum_{j=1}^{n-1}\frac{1}{l+2-j}\right]. & \quad n-l-2 \le 0
\end{cases}
\end{align}
Where the divergent part is encoded only in $J^1_1$ which can be calculated as
\begin{align}
J^1_1 = \frac{i}{(4\pi)^2}\left(\frac{2}{\epsilon} - \gamma -\log(m^2/4\pi)+1+\mathcal{O}(\epsilon)\right).
\end{align}
In the case where $l=n$, we have
\begin{align}
J_n^n=\frac{n(n+1)}{2}\left[J^1_1-\frac{i}{(4\pi)^2}\sum_{j=1}^{n-1}\frac{1}{n+2-j}\right].
\end{align}
In particular, the following integrals appear in the calculations which can be computed as
\begin{align}
\frac{1}{m^4}\int\frac{\text{d}^4 k}{(2\pi)^4}\ \frac{k^{2(l-1)}k_{\mu}k_{\nu}}{(k^2+m^2)^n} = \frac{1}{4}\delta_{\mu\nu} J_n^l,
\end{align}
and
\begin{align}
\frac{1}{m^4}\int\frac{\text{d}^4 k}{(2\pi)^4}\ \frac{k^{2(l-2)}k_{\mu}k_{\nu}k_{\rho}k_{\sigma}}{(k^2+m^2)^n} = \frac{1}{24}\delta_{\mu\nu\rho\sigma} J_n^l,
\end{align}
with
\begin{align}
\delta_{\mu\nu\rho\sigma} \equiv (\delta_{\mu\nu}\delta_{\rho\sigma}+\delta_{\mu\rho}\delta_{\nu\sigma}+\delta_{\mu\sigma}\delta_{\nu\rho}).
\end{align}
Also in this paper we have used the Feynman parametrization defined as
\begin{align}
\frac{1}{A_1^{\alpha_1}A_2^{\alpha_2}...A_n^{\alpha_n}} = \int_0^1 \text{d}x_1\int_0^1\text{d}x_2...\int_0^1\text{d}x_n\ \delta\left(\sum_{i=1}^n x_i-1\right)\frac{\prod_{i=1}^nx_i^{\alpha_i-1}}{\left(\sum_{i=1}^n x_iA_i\right)^{\sum_{i=1}^n \alpha_i}} \ \frac{\Gamma\left(\sum_{i=1}^n \alpha_i\right)}{\prod_{i=1}^n \Gamma\left(\alpha_i\right)}.
\end{align}

\section{The amplitudes for the 4-point functions}\label{app2}
The amplitude for the one-loop correction of 4-point function from the triangle diagram in figure \eqref{fig5} in the special case of equal momenta, $p_1=p_2=-p_3=-p_4=p$, can be written as
\begin{align}\label{apb1}
&\mathcal{M}_{\text{4pt,b; eq. mom.}} = 4! \cdot 3! \cdot  \ (-i) \epsilon_{r_1}^{\mu}(p)\epsilon_{r_2}^{\nu}(p)\epsilon_{r_3}^{\alpha*}(-p)\epsilon_{r_4}^{\beta*}(-p)\, J_1^1 \nonumber\\&\quad\times
\left[
\frac{
	3\delta_{\mu\nu} \left(
	p^2 \left(-15 m^4+125 m^2 p^2-327 p^4\right) \delta_{\alpha\beta}
	+10 \left(3 m^4-17 m^2 p^2+37 p^4\right) p_{\alpha} p_{\beta}
	\right)
}{720 M_2^2 M_4^4}\right.\nonumber\\&\quad\left.
+\frac{3p^2 \delta_{\alpha\nu} \left(15 m^4-40 m^2 p^2+71 p^4\right) \delta_{\beta\mu}
	+3p^2 \left(15 m^4-40 m^2 p^2+71 p^4\right) \delta_{\alpha\mu} \delta_{\beta\nu}}{720 M_2^2 M_4^4}\right.\nonumber\\ &\quad\left.
-\frac{3 p_{\nu} p^2 \left(443 p^2-140 m^2\right) \left(
	p_{\beta} \delta_{\alpha\mu}
	+p_{\alpha} \delta_{\beta\mu}
	\right)
	+3 p^2 \left(443 p^2-170 m^2\right) p_{\mu} \left(
	p_{\beta} \delta_{\alpha\nu}
	+p_{\alpha} \delta_{\beta\nu}
	\right)
}{360 M_2^2 M_3^2 M_4^2}\right.\nonumber\\ &\quad\left.
+\frac{
	p^2 p_{\nu} \left(
	p_{\mu} \left(
	\left(30 m^2+131 p^2\right) \delta _{\alpha\beta}
	+46 p_{\alpha} p_{\beta}
	\right)
	-3 \left(133 p^2-40 m^2\right) \left(
	p_{\beta} \delta_{\alpha\mu}+p_{\alpha} \delta_{\beta\mu}
	\right)
	\right)
}{720 M_2^2 M_4^4}\right.\nonumber\\ &\quad\left.
-\frac{p^2 \left(270 m^2+3181 p^2\right) p_{\mu} p_{\nu} \delta_{\alpha\beta}}{60 M_1^2 M_3^4}+\frac{p^2 \left(229 p^2-110 m^2\right) p_{\mu} p_{\nu} \delta_{\alpha\beta}}{20 M_1^2 M_3^2 M_4^2}\right.\nonumber\\ &\quad\left.
+\frac{p^2 \delta_{\alpha\beta} \left(p^2 \left(125 m^2-398 p^2\right) \delta_{\mu\nu}+2 \left(5 m^2+27 p^2\right) p_{\mu} p_{\nu}\right)}{120 M_1^2 M_4^4}\right.\nonumber\\ &\quad\left.
-\frac{p_{\mu} p_{\nu} \left(p^2 \left(15 m^2+3412 p^2\right) \delta_{\alpha\beta}+5 \left(150 m^2-821 p^2\right) p_{\alpha} p_{\beta}\right)}{180 M_2^2 M_3^4}\right.\nonumber\\ &\quad\left.
+\frac{
	p_{\nu} p_{\mu} \left(
	p^2 \left(2287 p^2-930 m^2\right) \delta_{\alpha\beta}
	+10 \left(29 p^2-12 m^2\right) p_{\alpha} p_{\beta}
	\right)}{360 M_2^2 M_3^2 M_4^2}\right.\nonumber\\ &\quad\left.
+\frac{
	6 p^2 \left(46 p^2-25 m^2\right) p_{\alpha} p_{\mu } \delta_{\beta\nu}
	+6p^2 \delta_{\alpha\nu} \left(46 p^2-25 m^2\right) p_{\beta} p_{\mu}}{720 M_2^2 M_4^4}\right]
\end{align}
Besides that, the amplitude for the one-loop corrections to the 4-point function of the generalized Proca theory through the square diagram in figure \eqref{fig5} in the special case of equal momenta, $p_1=p_2=-p_3=-p_4=p$, can be simplified as
\begin{align}\label{apb2}
\mathcal{M}&_{\text{4pt,c; eq.mom.}} = 4! \cdot 4! \cdot (-i)\,\epsilon^\alpha_{r_1}(p)\epsilon^\xi_{r_2}(p)\epsilon^{\rho*}_{r_3}(-p)\epsilon^{\sigma*}_{r_4}(-p) J^1_1 \nonumber\\&\times\left\{
\frac{1}{12 M_3^8}\left(144 m^4+2592 m^2 p^2+28961 p^4\right) p_{\alpha} p_{\zeta} p_{\rho} p_{\sigma}\nonumber\right.\\&\left.
+\frac{1}{3 M_3^6 M_4^2}\left(36 m^4+639 m^2 p^2-2377 p^4\right) p_{\alpha} p_{\zeta} p_{\rho} p_{\sigma}\nonumber\right.\\&\left.
+\frac{p^4}{288 M_4^8} \left[
27 m^4 \delta_{\alpha\sigma} \delta_{\zeta\rho}
+27 m^4 \delta_{\alpha\rho} \delta_{\zeta\sigma}
-414 m^2 p^2 \delta_{\alpha\sigma} \delta_{\zeta\rho}
-414 m^2 p^2 \delta_{\alpha\rho} \delta_{\zeta\sigma}
\nonumber\right.\right.\\&\left.\left.
+2 p_{\rho} \left(
\left(331 p^2-36 m^2\right) p_{\sigma} \delta_{\alpha\zeta}
+2 \left(9 m^2-52 p^2\right) p_{\zeta} \delta_{\alpha\sigma}
\right)
\nonumber\right.\right.\\&\left.\left.
+2 p_{\alpha} \left(
2 \left(9 m^2-52 p^2\right) \left(
p_{\sigma} \delta_{\zeta\rho}
+p_{\rho} \delta_{\zeta\sigma}
\right)
+p_{\zeta} \left(
\left(109 p^2-36 m^2\right) \delta_{\rho\sigma}
+48 p_{\rho} p_{\sigma}
\right)
\right)
\nonumber\right.\right.\\&\left.\left.
+36 m^2 p_{\zeta} p_{\sigma} \delta_{\alpha\rho}
+3 \left(9 m^4-138 m^2 p^2+593 p^4\right) \delta_{\alpha\zeta} \delta_{\rho\sigma}
+1779 p^4 \delta_{\alpha\sigma} \delta_{\zeta\rho}
+1779 p^4 \delta_{\alpha\rho} \delta_{\zeta\sigma}
\nonumber\right.\right.\\&\left.\left.
-208 p^2 p_{\zeta} p_{\sigma} \delta_{\alpha\rho}
\right]\nonumber\right.\\&\left.
-\frac{p^2}{144 \text{M3}^2 \text{M4}^6}\left[
\left(54 m^4-792 m^2 p^2+2597 p^4\right) p_{\zeta} p_{\sigma} \delta_{\alpha\rho}
\nonumber\right.\right.\\&\left.\left.
+p_{\rho} \left(
\left(54 m^4-432 m^2 p^2-1663 p^4\right) p_{\sigma} \delta_{\alpha\zeta}
+\left(54 m^4-792 m^2 p^2+2597 p^4\right) p_{\zeta} \delta_{\alpha\sigma}
\right)
\nonumber\right.\right.\\&\left.\left.
+p_{\alpha} \left(
\left(54 m^4-792 m^2 p^2+2597 p^4\right) \left(
p_{\sigma} \delta_{\zeta\rho}
+p_{\rho} \delta_{\zeta\sigma}
\right)
+p_{\zeta} \left(
\left(54 m^4-1152 m^2 p^2+6029 p^4\right)\delta_{\rho\sigma}
\right.\right.\right.\right.\nonumber\\&\left.\left.\left.\left.-12 \left(18 m^2+43 p^2\right) p_{\rho} p_{\sigma}
\right)
\right)
\right]\nonumber\right.\\&\left.
+\frac{1}{144 M_3^4 M_4^4}\left[
p^2 \left(
\left(
-72 m^4-1638 m^2 p^2
+21337 p^4
\right) p_{\zeta} p_{\sigma} \delta_{\alpha\rho}
\nonumber\right.\right.\right.\\&\left.\left.\left.
+p_{\rho} \left(
\left(
-72 m^4-1242 m^2 p^2
+21997 p^4
\right) p_{\sigma} \delta_{\alpha\zeta}
+\left(
-72 m^4-1638 m^2 p^2
+21337 p^4
\right) p_{\zeta} \delta_{\alpha\sigma}
\right)
\right)
\nonumber\right.\right.\\&\left.\left.
+p_{\alpha} \left(
p^2 \left(
-72 m^4-1638 m^2 p^2+21337 p^4
\right) \left(
p_{\sigma} \delta_{\zeta\rho}+p_{\rho} \delta_{\zeta\sigma}
\right)
\nonumber\right.\right.\right.\\&\left.\left.\left.
+p_{\zeta} \left(p^2
\left(
-72 m^4 -1962 m^2 p^2 +21469 p^4
\right) \delta_{\rho\sigma}
+12 \left(
36 m^4 +156 m^2 p^2 +193 p^4
\right) p_{\rho} p_{\sigma}
\right)
\right)
\right]
\phantom{\frac{1}{M_3^2}}\hspace{-0.74cm}\right\}.
\end{align}

\end{document}